\documentclass[9pt,twocolumn,twoside]{opticajnl}
\journal{opticajournal} 

\setboolean{shortarticle}{false}

\usepackage{amsmath,amssymb,amsfonts}
\usepackage{graphicx}
\usepackage{textcomp}
\usepackage{xcolor}

\usepackage{multirow}
\usepackage{booktabs}
\usepackage{array}
\usepackage{url}
\usepackage{tikz}
\usetikzlibrary{positioning, decorations.pathreplacing, calligraphy, shapes.geometric}

\usepackage{siunitx}

\title{Transition-Aware Routing in Hybrid Hollow-Core/Single-Mode Fiber Networks: A Cost--Throughput Investigation}

\author[1,*]{Md Ghulam Saber}
\author[]{Zhiping Jiang}

\affil[]{Ottawa Research Center, Huawei Technologies Canada, 303 Terry Fox Drive, Kanata, ON, K2K 3J1, Canada.}

\affil[*]{md.ghulam.saber@huawei.com}

\begin{abstract}
Incremental deployment of hollow-core fiber (HCF) in single-mode-fiber
(SMF) networks introduces a routing tradeoff: reducing HCF--SMF
transitions can improve physical-layer feasibility, but overly
transition-averse routing incurs harmful path detours. We study this
tradeoff with a common event-driven simulator comparing six
protected routing schemes spanning fiber-blind, generalized
signal-to-noise ratio (GSNR)-aware, and explicitly transition-aware
designs on hybrid HCF/SMF topologies, using a per-transition GSNR
penalty and an exploratory splice-failure availability term. Across
six reference topologies, five HCF deployment fractions, and dynamic
loads at 300~Erlang, the strongest transition
minimizers---transition-penalty-aware routing (TPAR) and the
GSNR/fiber-transition joint scheme (GFJ)---halve the mean transition
count at a 20--25\% carried-traffic penalty. Among the intermediate
designs, GSNR-maximal routing with transition-aware reranking (GMR-T)
cuts transitions by $\sim$22\% versus distance-adaptive routing and
spectrum assignment (DA-RSA) at a 3\% throughput cost, and
bounded-detour TPAR (BD-TPAR) by $\sim$11\% at only a 1\% cost. Deployment
pattern also matters: contiguous HCF rollout lowers transitions by
$\sim$40\% on average while improving carried traffic, reducing the
benefit of aggressive transition-aware routing. These results support BD-TPAR as
a practical default under fragmented deployment, GMR-T as a
lower-complexity alternative, and TPAR/GFJ only where external
transition cost is high.
\end{abstract}

\setboolean{displaycopyright}{false} 

\doi{} 

\begin{document}

\maketitle

\section{Introduction}\label{sec:intro}
\vspace{-.35cm}
The gradual transition from all-SMF optical networks to hybrid HCF/SMF infrastructures introduces a routing cost that conventional optical-network models do not capture. That cost is not the ordinary in-span fusion splice---low-loss and already in every network's per-kilometer loss budget---but the fiber-type transition itself. Each HCF$\leftrightarrow$SMF interface is a mode-field-mismatch junction needing a specialized coupler with higher, more variable loss, and is simultaneously a discontinuity in chromatic dispersion (CD), group velocity, span-loss profile, and nonlinearity that forces the downstream erbium-doped fiber amplifier (EDFA) to retarget its gain~\cite{saber2026ecoc}. These costs arise only where fiber type changes, so they are absent in all-SMF or all-HCF backbones and grow with the number of transitions a route crosses. A route attractive under length or GSNR alone may thus be undesirable once transitions are counted, while avoiding them too aggressively forces detours that lengthen paths, cut GSNR margin, and hurt carried traffic. The design question is therefore not whether transition awareness helps, but when the reduction in cross-fiber transitions is worth the routing cost it introduces.

This question sits at the intersection of two established lines of work. The first is impairment-aware routing and spectrum assignment (RSA), where shortest-path, disjoint-path, and quality-of-transmission (QoT)-aware provisioning methods are commonly built on Yen-style K-shortest paths, Suurballe-type protection, and Gaussian-noise (GN) model-driven GSNR estimation \cite{yen1971finding,suurballe1974disjoint,chatterjee2015rsa,ramamurthy2003survivable,azodolmolky2009impairment,poggiolini2012gnmodel,gnpy2020,ferrari2020gnpy,jinno2010distance}. These methods are mature and effective, but they usually use physical length or inverse GSNR as the routing weight, which leaves fiber type invisible during path selection. The second line is the emerging literature on HCF systems and networks. HCF has matured toward transport-grade deployment: recent antiresonant designs reach attenuation at or below the silica Rayleigh floor---0.091~dB/km, and below 0.2~dB/km across a 66~THz window~\cite{petrovich2025broadband}---while preserving HCF's low latency and extremely weak Kerr nonlinearity \cite{petrovich2025broadband,ofc2026hcf,Zhu2020HCFCableLowLatency,Nespola2021HCF618km,Hong2025BeyondTbHCF,Gao2025HCF100kmCL}. At the same time, the relevant penalties are now better understood, including intermodal interference (IMI), HCF-to-SMF splice loss, and $CO_2$ absorption in L-band operation \cite{poggiolini2022hcf,photonics13060559,NumkamFokoua2023_AOP_LossReview,VanNewkirk2016_OL_S2_Antiresonant,Nespola2021_OFC_F3B5_ReducedIMI,Ge2023_OFC_W4D6_KerrHCF,shi2024splicing,Suslov2022_OE_BackReflection,wang2025co2,saber2026_hcfimdd,Li2025_ACP_LowGasAbsorptionHCF}. 

What remains less developed is the routing-decision layer for dynamic protected provisioning in hybrid topologies. Prior work has examined HCF placement, hybrid HCF/SMF planning, and joint fiber, modulation, and spectrum allocation, and a companion study has considered protection switching on the same topology family \cite{Sticca2025CapacityScaling,Correia2026_Photonics_HCF_SystemSpecs,Ouyang2025HybridULL,saber2026ecoc}. Dynamic impairment-aware provisioning itself is mature---ABACUS jointly optimizes routing, modulation, and spectrum under a QoT constraint~\cite{kiran2024abacus}, and Ouyang {et al.} allocate routing, fiber, modulation, and spectrum for hybrid ultra-low-loss/standard-SMF links~\cite{Ouyang2025HybridULL}. But their objectives are either fiber-type-blind or fix fiber type at planning time (which link gets which fiber) rather than pricing a per-route {transition} cost during path selection. To our knowledge, no prior work prices HCF$\leftrightarrow$SMF transitions explicitly in the routing objective for dynamic, dedicated-protection provisioning, and the practically important middle ground between ignoring transitions and minimizing them aggressively is likewise unexplored, especially where the operator wants some transition awareness but not arbitrary detours.

This paper addresses that gap with a controlled comparison of six protected routing schemes on a common event-driven simulator, spanning fiber-blind and GSNR-aware baselines through to explicitly transition-aware designs. To make transitions visible without abandoning the GN-model framework, a per-transition GSNR penalty---calibrated from HCF--SMF splice loss and the EDFA gain transients induced at reconfigurable optical add/drop multiplexers (ROADMs)---is applied at the feasibility stage, symmetrically across all schemes, so even a fiber-blind objective pays the physical-layer cost of transitions at provisioning time. A composite per-demand availability metric with an exploratory splice-related failure term, computed with the ITU-T G.911 steady-state methodology (its failure rates from field cable-cut statistics and a conservative splice assumption), serves as a reliability cross-check rather than an optimization target. Two middle-ground constructions are introduced to fill the gap between familiar baselines and aggressive transition minimization: GMR-T keeps the candidate set of GSNR-maximal routing (GMR) but reranks feasible candidates by transition count, while BD-TPAR scheme retains the transition awareness of the TPAR scheme but caps the detour length and falls back to GMR when no capped candidate exists. The paper thus (i)~compares six schemes under a common transition-penalized feasibility model, (ii)~introduces two intermediate designs that make transition awareness practical, and (iii)~converts the trade-off into a decision rule mapping operator cost structure to scheme choice. The results show that the preferred scheme depends on deployment pattern and cost regime rather than on a universal ranking: BD-TPAR emerges as the practical default under fragmented rollout, GMR-T as a lower-complexity alternative, DA-RSA or GMR as sufficient when no explicit transition cost is carried, and TPAR or the GFJ scheme as justified only when transition cost is high enough to warrant the extra detour.

Section~\ref{sec:system} describes the model; Section~\ref{sec:schemes} the six schemes; Section~\ref{sec:sens} the parameters; Section~\ref{sec:results_c} the results; and Section~\ref{sec:discussion} the decision rule.
\section{System Model}\label{sec:system}

 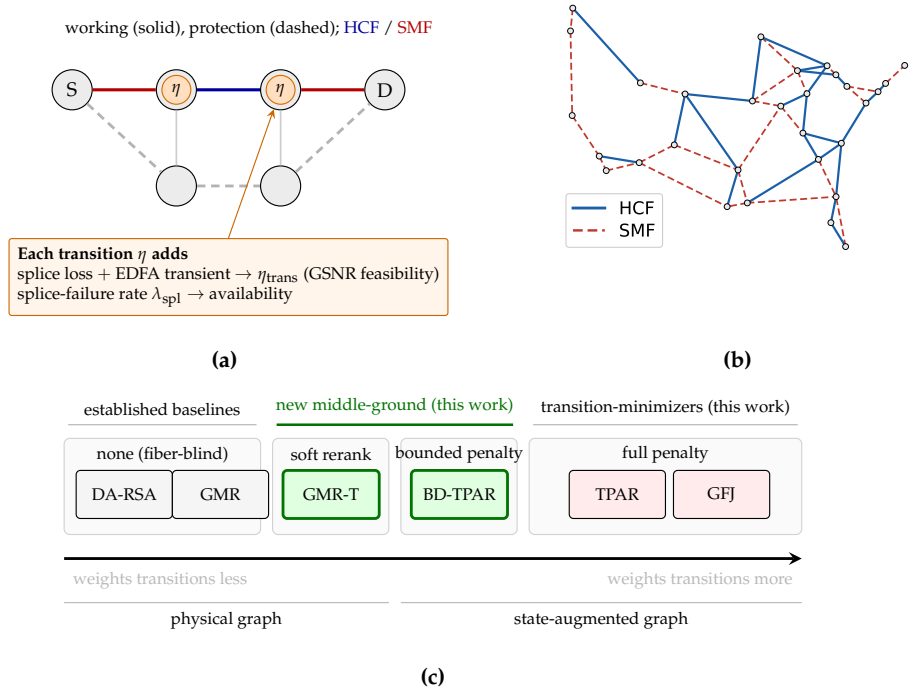
\begin{figure*}[t]
	\centering
	\scalebox{1.06}{%
	\begin{tikzpicture}[
		font=\footnotesize,>=stealth,
		netnode/.style={circle,draw,fill=gray!15,minimum size=5mm,inner sep=0pt},
		hcf/.style={line width=1.1pt,blue!60!black},
		smf/.style={line width=1.1pt,red!70!black},
		prot/.style={line width=1pt,gray!60,densely dashed},
		mesh/.style={line width=0.6pt,gray!40},
		eta/.style={circle,draw=orange!85!black,fill=orange!25,minimum size=3.6mm,
			inner sep=0pt,font=\scriptsize},
		cost/.style={draw=orange!80!black,fill=orange!8,rounded corners=1pt,
			align=left,inner sep=3pt,font=\scriptsize},
		sbox/.style={rounded corners=1.5pt,draw,minimum width=12mm,minimum height=6mm,
			align=center,inner sep=1.5pt,font=\scriptsize},
		base/.style={sbox,fill=gray!8},
		mid/.style={sbox,draw=green!45!black,line width=1pt,fill=green!12},
		aggr/.style={sbox,fill=red!8},
		zone/.style={rounded corners=2pt,draw=gray!35,fill=gray!4}]
		
		\node[netnode](S)  at (0,0.7)   {S};
		\node[netnode](n1) at (1.3,0.7) {};
		\node[netnode](n2) at (2.6,0.7) {};
		\node[netnode](D)  at (3.9,0.7) {D};
		\node[netnode](m1) at (1.3,-0.5){};
		\node[netnode](m2) at (2.6,-0.5){};
		\draw[smf](S)--(n1); \draw[hcf](n1)--(n2); \draw[smf](n2)--(D);
		\draw[prot](S)--(m1); \draw[prot](m1)--(m2); \draw[prot](m2)--(D);
		\draw[mesh](n1)--(m1); \draw[mesh](n2)--(m2);
		\node[eta] at (n1){$\eta$}; \node[eta] at (n2){$\eta$};
		\node[font=\scriptsize,anchor=west] at (-0.2,1.45)
		{working (solid), protection (dashed); \textcolor{blue!60!black}{HCF} / \textcolor{red!70!black}{SMF}};
		\node[cost,anchor=north] at (1.95,-1.15) {%
			\textbf{Each transition $\eta$ adds}\\
			splice loss $+$ EDFA transient $\to \eta_\text{trans}$ (GSNR feasibility)\\
			splice-failure rate $\lambda_\text{spl}\to$ availability};
		\draw[->,orange!80!black] (1.95,-1.15) -- (n2);
		\node[font=\small\bfseries,anchor=north] at (1.9,-2.45) {(a)};

		\node[anchor=north] at (8.3,2.05)
		{\includegraphics[width=4.7cm]{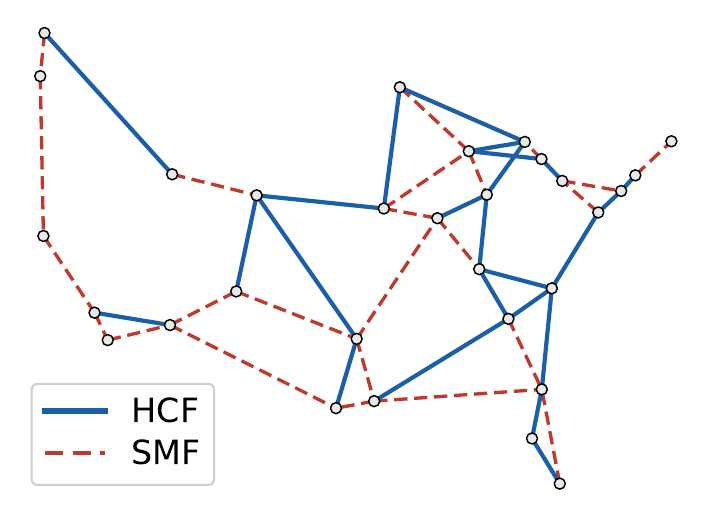}};
		\node[anchor=north,font=\small\bfseries] at (8.3,-2.45) {(b)};

		\begin{scope}[shift={(0,-4.7)}]
			\node[font=\small\bfseries,anchor=north] at (4.5,-1.7) {(c)};
			\fill[zone] (-0.1,-0.15) rectangle (2.35,1.05);
			\fill[zone] (2.5,-0.15)  rectangle (3.95,1.05);
			\fill[zone] (4.1,-0.15)  rectangle (5.55,1.05);
			\fill[zone] (5.7,-0.15)  rectangle (9.1,1.05);
			\node[font=\scriptsize] at (1.13,0.85){none (fiber-blind)};
			\node[font=\scriptsize] at (3.23,0.85){soft rerank};
			\node[font=\scriptsize] at (4.83,0.85){bounded penalty};
			\node[font=\scriptsize] at (7.40,0.85){full penalty};
			\node[base] at (0.65,0.35){DA-RSA};
			\node[base] at (1.85,0.35){GMR};
			\node[mid]  at (3.23,0.35){GMR-T};
			\node[mid]  at (4.83,0.35){BD-TPAR};
			\node[aggr] at (6.80,0.35){TPAR};
			\node[aggr] at (8.10,0.35){GFJ};
			\draw[->,line width=1pt] (-0.1,-0.45) -- (9.1,-0.45);
			\node[anchor=north west,font=\scriptsize,text=gray!55] at (-0.1,-0.5){weights transitions less};
			\node[anchor=north east,font=\scriptsize,text=gray!55] at (9.1,-0.5){weights transitions more};
			\draw[gray!55] (-0.1,-1.0) -- node[below,font=\scriptsize,text=black]{physical graph} (3.95,-1.0);
			\draw[gray!55] (4.1,-1.0)  -- node[below,font=\scriptsize,text=black]{state-augmented graph} (9.1,-1.0);
			\draw[gray!60] (-0.05,1.22)
			-- node[above,font=\scriptsize,text=black]{established baselines} (2.30,1.22);
			\draw[green!45!black,line width=0.9pt] (2.55,1.22)
			-- node[above,font=\scriptsize,text=green!45!black]{new middle-ground (this work)} (5.50,1.22);
			\draw[gray!60] (5.75,1.22)
			-- node[above,font=\scriptsize,text=black]{transition-minimizers (this work)} (9.05,1.22);
		\end{scope}
	\end{tikzpicture}%
	}
	\vspace{-.3cm}
	\caption{Overview. (a) A 1+1-protected demand traverses HCF and SMF segments; each
		HCF$\leftrightarrow$SMF transition ($\eta$) adds splice loss and an EDFA gain transient
		to the Extended-GSNR feasibility budget [\eqref{eq:eta_trans}] and an exploratory
		splice-failure term to the availability model (Section~\ref{sec:availability}). (b) CORONET fiber-type map at 50\% HCF under
		uniform-random deployment (one seed); the interleaving of HCF (blue solid) and SMF
		(red dashed) drives the high per-route transition counts. (c) The six schemes
		organized by how strongly path selection weights transitions: fiber-blind baselines
		(DA-RSA, GMR), the middle-ground schemes introduced here (GMR-T, BD-TPAR), and the
		explicit minimizers (TPAR, GFJ).}
		\vspace{-.3cm}
	\label{fig:overview}
\end{figure*}

\subsection{Topology ensemble and fiber assignment}\label{sec:topo}
We use six published reference topologies: CORONET (30),
COST239 (11), NSFNET (14), USNET (24), and COST266 (37) and
Nobel-Germany (17). The HCF
deployment fraction is swept over $p_\text{HCF}\in\{0,0.25,0.5,0.75,1.0\}$; the default
assignment marks a uniformly random edge subset as HCF until the count reaches
$p_\text{HCF}\,|E|$ (requiring no operator coordination), and as an alternative we also
consider a contiguous breadth-first-search (BFS)-expanded HCF cluster grown from a random
seed link (Section~\ref{sec:rollout_compare}). For each
$(\text{topology},p_\text{HCF},\text{seed})$ triplet, all six schemes see the same
deterministic HCF link set and the same random variates---identical
source--destination pairs and holding-time draws from a fixed seed---so scheme comparison
is paired by construction. Across loads the arrival {rate} is scaled by the Erlang
value while these variates are held fixed, so the load sweeps vary offered load without
changing the underlying demand realization.

\paragraph{How the hybrid plant is implemented.} Fiber type is
assigned at {link} granularity: a link is all-HCF or all-SMF
over its whole length and is operated as a chain of single-type
amplified spans (80~km for SMF, 100~km for antiresonant HCF
(AR-HCF); Table~\ref{tab:fiber_params}), with amplified spontaneous
emission (ASE) noise and nonlinear interference (NLI) accumulated
under per-fiber locally-optimized (LOGO) launch power,
capped so the full-fill aggregate
stays within a \SI{+23}{dBm} booster ceiling. Since LOGO
makes the per-span GSNR invariant along a uniform-fiber link, we
accumulate inverse-GSNR by scaling one reference span's value by the
link length rather than summing spans individually. No link mixes fiber types, so
HCF$\leftrightarrow$SMF interfaces occur only at ROADM nodes where a
route changes fiber type; $N_\text{trans}(P)$ counts these
handovers, and each link's latency follows its own group index.

\subsection{Extended GN model with IMI}\label{sec:extgn}
The per-link GSNR is the incoherent GN-model accumulation~\cite{gnpy2020}
\begin{equation}
	\frac{1}{\mathrm{GSNR}_\mathrm{ext}}=\frac{1}{\mathrm{GSNR}_\mathrm{ASE}}
	+\frac{1}{\mathrm{GSNR}_\mathrm{NLI}}+\frac{1}{\mathrm{GSNR}_\mathrm{IMI}},
	\label{eq:extgn}
\end{equation}
with standard ASE and NLI terms; the IMI term is non-zero only on HCF links and scales
linearly with length. CO\textsubscript{2} absorption on L-band HCF enters through the
wavelength-dependent loss $\alpha(\lambda)$ (Section~\ref{sec:intro}), hence through the
ASE term. The NLI term grows with the number of lit channels. To keep routing fast we do
not recompute the GSNR on every arrival: each link's occupancy is classified as
\textsc{low}, \textsc{med}, or \textsc{high} (breakpoints near one third and two thirds
of the 80 channels, with a small hysteresis band), the GSNR is precomputed once per
state, and a link's routing weight changes only when its occupancy class changes.
Figure~\ref{fig:overview} sketches the hybrid-path structure and the design space of the
six schemes, and Table~\ref{tab:fiber_params} lists the fixed fiber parameters.

\begin{table}[t]
	\centering
	\caption{Fiber parameters used in the simulation. The HCF values
	are a conservative scenario---recent demonstrations report lower
	attenuation ($\sim$0.09~dB/km) and IMI---so the HCF advantage here
	is a lower bound.}
	\label{tab:fiber_params}
	\small
	\begin{tabular}{lcc}
		\toprule
		Parameter (C / L) & SMF (G.652) & AR-HCF \\
		\midrule
		$\alpha$ [dB/km]                 & 0.20 / 0.21 & 0.13 / 0.14 \\
		$D$ [ps/(nm$\cdot$km)]           & 17.0 / 20.0 & 3.5 / 4.0 \\
		$\gamma$ [W$^{-1}$km$^{-1}$]     & 1.3 / 1.15  & $10^{-3}$ \\
		$n_g$                            & 1.468       & 1.0003 \\
		Span length [km]                 & 80          & 100 \\
		EDFA NF (C / L) [dB]             & 5.5 / 6.5   & 5.5 / 6.5 \\
		IMI MPI level [dB/km]            & ---         & $-55$ / $-53$ \\
		Pulse roll-off $\alpha_\text{RRC}$ & 0.10      & 0.10 \\
		\bottomrule
	\end{tabular}
\end{table}

\subsection{Per-transition GSNR penalty}\label{sec:eta_trans}
Each HCF$\leftrightarrow$SMF interface costs the physical layer an SNR budget independent
of the routing weight, which we model as a per-transition penalty $\eta_\text{trans}$
calibrated from two literature-informed contributions. First, HCF--SMF splice/coupler loss, which recent work has
reduced from about \SI{1.2}{\decibel} (early angle-cleaved splices) to \SI{0.15}{\decibel}
on antiresonant nodeless designs~\cite{shi2024splicing}; we adopt \SI{0.3}{\decibel} as a
conservative mid-range value. Second, the ROADM-induced EDFA gain
transient: cascaded EDFAs in colorless--directionless--contentionless (CDC) ROADMs react to add/drop events with surviving-channel
excursions of several decibels that gain-control loops must
suppress~\cite{srivastava1997edfatancevski1999swings}, and
residual gain-ripple and filter-shape uncertainty is a recognized QoT-margin
component~\cite{mahajan2020edfa}; we attribute a conservative \SI{0.1}{\decibel} per
transition to gain re-targeting at each fiber-type boundary. This second term is an
order-of-magnitude modeling estimate, not a measured value: the cited work establishes
that such transients exist and consume margin, but to our knowledge no study reports a
single calibrated per-transition GSNR figure. We therefore treat
$\eta_\text{trans}=\SI{0.4}{\decibel}$ as a nominal operating point whose influence on the
ranking is bounded by the sweep of Section~\ref{sec:eta_sweep}. It is applied symmetrically
(splice loss is reciprocal)
and subtracted from the path-level Extended-GSNR (in dB),
\begin{equation}
	\mathrm{GSNR}_\mathrm{ext}^\text{path}=
	-10\log_{10}\!\Big(\textstyle\sum_{i\in P}1/\mathrm{GSNR}_\mathrm{ext}^{(i)}\Big)
	-N_\text{trans}(P)\,\eta_\text{trans},
	\label{eq:eta_trans}
\end{equation}
where $N_\text{trans}(P)$ counts transitions on path $P$. The penalty is charged at the
feasibility test on {both} paths of a 1+1 demand (each must clear its threshold
post-switchover), but the transition-aware objectives and the reported per-demand count
$\bar{n}$ (Fig.~\ref{fig:mod_lift}(b)) score the {working} path only---the path
carrying traffic in normal operation. Being a feasibility test rather than a routing
objective, the penalty is paid by every scheme regardless of whether its path-selection
objective weights $N_\text{trans}$.
Digital-signal-processing (DSP) equalizer reconvergence is deliberately excluded: modern
1+1 deployments run both paths in hot standby with pre-converged DSP, so switchover is electrical
selection~\cite{tang2024multilayer} and that cost surfaces as transponder resource use
rather than per-channel GSNR. Throughout, $\eta_\text{trans}=\SI{0.4}{\decibel}$ should be
read as a calibrated sensitivity parameter that sets the operating point of the study, not
as a measured universal constant; Section~\ref{sec:eta_sweep} sweeps it over $[0,1]$~dB
and shows that the scheme ranking is insensitive to the calibration.

\subsection{Provisioning constraints and dynamic traffic}\label{sec:mod}
Four dual-polarization (DP) formats \{DP-QPSK, DP-8QAM, DP-16QAM, DP-64QAM\} carry GSNR thresholds
$\theta_\text{mod}\in\{10,13,16,22\}$~dB; provisioning requires
$\mathrm{GSNR}_\mathrm{ext}^\text{path}\ge\theta_\text{mod}+m$ with a \SI{1}{\decibel}
margin $m$, where $\mathrm{GSNR}_\mathrm{ext}^\text{path}$ is the end-to-end
(path-level) GSNR of \eqref{eq:eta_trans}---the incoherent sum of per-link inverse-GSNR
less the transition penalty---not a per-link threshold. The selector picks the
highest-spectral-efficiency format that passes. For a 1+1 demand the binding constraint is
the worse of the working and protection path-level GSNRs.
The study band comprises 80 channels on a 75~GHz grid (64~GBd) spanning
191.307--197.232~THz (1520.0--1567.1~nm), a 6~THz window that extends past the
conventional C-band (1530--1565~nm) on the short-wavelength side, in the spirit of
extended/super-C-band systems; we retain the label ``C-band'' for brevity. The grid is
wavelength-continuity-constrained with no converters; first-fit assignment
ORs (bitwise logical) the per-link occupancy bitmaps of the working/protection pair and takes the
lowest free index. C-band CO\textsubscript{2} absorption is an order of magnitude weaker than in the
L-band~\cite{chen2025co2} and is not modeled.
The permitted working-versus-protection
latency asymmetry is bounded per topology by
\begin{equation}
	\Delta\tau_\mathrm{max}=\beta\,\bar{L}_\mathrm{sp}\,
	\left|\frac{1}{v_{g,\mathrm{SMF}}}-\frac{1}{v_{g,\mathrm{HCF}}}\right|,
	\label{eq:dtmax}
\end{equation}
with $\bar{L}_\mathrm{sp}$ the mean shortest-path length, $\beta=0.60$, and
$v_{g,f}=c/n_g^{(f)}$; the bound is enforced as a hard constraint (HC1). It exists
because the receiver must realign the two continuously arriving 1+1 copies within a
finite differential-delay buffer for switchover to be hitless. In all-SMF networks the
asymmetry comes from length differences alone and is rarely binding; in a hybrid plant
the $\sim$1.6~\si{\micro\second}/km group-delay gap between HCF and
SMF~\cite{saber2026ecoc} makes it first-order. \eqref{eq:dtmax} scales the allowance to
the topology: the group-delay term is the asymmetry a mean-length route would accumulate
if one path were all-SMF and the other all-HCF, and $\beta=0.60$ grants 60\% of that
worst case as buffer budget. Each trial issues $N=3{,}000$ requests (600 warm-up, 2400
measurement) with Poisson arrivals at rate $\rho/H$ ($\rho$ in Erlangs, mean holding
$H=\SI{1}{hour}$) and uniform source--destination pairs; every operating point runs ten
seeds, and all figures report 95\% confidence intervals over the seeds.

\subsection{Composite per-demand availability}\label{sec:availability}
Each accepted 1+1 demand is scored with the standard repairable series/parallel
model~\cite{tornatore2005availability}: the working path is a chain of $N_\text{L}^{(w)}$
cable spans and $N_\text{trans}^{(w)}$ splice joints (likewise the protection path), and
the demand is up if and only if either path is up. With per-link cable and per-splice availabilities, each set by a failure rate $\lambda$ and a mean time to repair (MTTR),
\begin{equation}
	a_\text{cab}=\frac{1}{1+\lambda_\text{cab}\,\mathrm{MTTR}_\text{cab}},\qquad
	a_\text{spl}=\frac{1}{1+\lambda_\text{spl}\,\mathrm{MTTR}_\text{spl}},
\end{equation}
we adopt $\mathrm{MTTR}_\text{cab}=\SI{12}{\hour}$---the Bellcore cable-cut repair time
tabulated by To and Neusy~\cite{toneusy1994}---and a deliberately conservative cable-cut
rate $\lambda_\text{cab}=\SI{1e-4}{\per\hour}$, of the same order as the reported field
statistic of 4.39 cable cuts per 1000 sheath-miles per year~\cite{toneusy1994}
($\sim\SI{3e-5}{\per\hour}$ per 80--100~km span); the steady-state form itself follows the
ITU-T~G.911 methodology~\cite{itu_g911}. The splice term
deserves a caveat: production fusion splices are normally treated as passive and
effectively failure-free in availability planning, and no per-splice outage rate is
standardized. We include splice reliability as a conservative what-if, since ITU-T G.911 lists splices and connectors as fiber-plant reliability parameters~\cite{itu_g911}. We assume $\lambda_\text{spl}=\SI{1e-5}{\per\hour}$ and $\mathrm{MTTR}_\text{spl}=\SI{6}{\hour}$, with a failure rate ten times lower than that of the cable. 
 The purpose is to test whether transition-aware routing could {ever} earn a
reliability benefit; Section~\ref{sec:results_avail} shows the answer is negative at
any plausible rate, so this choice drives no conclusion.
Per-path availability is the series product
$A^{(w)}=a_\text{cab}^{\,N_\text{L}^{(w)}}\,a_\text{spl}^{\,N_\text{trans}^{(w)}}$, and the
1+1 composite adds a coarse channel-outage term keyed to a single binary label (whether
the working path is majority-HCF)---a first-order model, not a per-link fiber-aware
reliability calculation, meant only to bound whether transition count could reorder the
schemes on availability. Only the splice term depends on $N_\text{trans}$; its impact is
swept in Section~\ref{sec:results_avail}.

\section{Routing Schemes}\label{sec:schemes}

All six schemes run on a common provisioning loop
(Section~\ref{sec:loop}) and all six pay the per-transition GSNR
penalty of \eqref{eq:eta_trans} at the feasibility check; they
differ only in how strongly path {selection} weights
transitions, from not at all (DA-RSA, GMR), through the soft
middle-ground designs introduced here (GMR-T, BD-TPAR), to explicit
minimization on a state-augmented graph (TPAR, GFJ), as organized in
Fig.~\ref{fig:overview}(c).

\subsection{DA-RSA: distance-adaptive RSA}\label{sec:darsa}
The DA-RSA baseline performs Yen's K-shortest-paths (K-SP)
search~\cite{yen1971finding} on physical link length and uses the
Extended-GSNR feasibility check (\eqref{eq:eta_trans}, including
the per-transition penalty) to discard infeasible candidates. The
highest-spectral-efficiency modulation format passing the
threshold-plus-margin inequality is selected, and among feasible
candidates the shortest-length path is chosen. We use $K=5$
candidate paths per demand, for all schemes. QoT-aware RSA studies
typically use $K\in[3,10]$~\cite{chatterjee2015rsa,gnpy2020}: below
$\sim$3 sparse topologies often lack a feasible working/protection
pair; beyond $\sim$5 the extra candidates are longer variants that
rarely pass the GSNR gate while runtime grows linearly in $K$. $K=5$
is the midpoint, held fixed across schemes for fairness. DA-RSA does
not weight HCF$\leftrightarrow$SMF transitions in its objective,
but transitions still drive its blocking probability by reducing the
post-penalty GSNR of transition-heavy candidates below the
modulation threshold.

\subsection{GMR: GSNR-maximal routing}\label{sec:gmr}
GMR uses the per-link inverse Extended-GSNR as a Dijkstra edge
weight,
$w_{ij}^\text{GMR} = 1/\mathrm{GSNR}_\text{ext}^{(ij)}$.
Summing along a path recovers the inverse end-to-end GSNR before the
transition penalty, so the shortest-weight path is the
highest-pre-penalty-GSNR path. The post-routing feasibility check
applies \eqref{eq:eta_trans} -- meaning that, like DA-RSA, GMR
sees transitions only at the feasibility gate. GSNR-shortest paths
nevertheless tend to favor HCF (higher per-link GSNR), which is
transition-light, giving GMR an implicit advantage on hybrid
topologies (Section~\ref{sec:results_c}).

\subsection{GMR-T: GMR with K-best transition reranking}\label{sec:gmr_t}
GMR-T uses the same candidate generator as GMR -- Yen's
K-shortest-paths on the per-link inverse-Extended-GSNR -- but
deviates from GMR in the selection step. Instead of taking the
GMR-first feasible candidate, GMR-T enumerates the entire $K$-best
set, applies the same feasibility checks (HC1--HC3), and selects
the feasible candidate with the lowest cross-fiber transition
count, breaking ties by GMR cost. Concretely, among the feasible
set $\mathcal{F}\subseteq\{1,\ldots,K\}$, GMR-T picks
\begin{equation}
	k^\star = \arg\min_{k\in\mathcal{F}}
	\bigl(N_\text{trans}(P_k),
	-\mathrm{GSNR}_\text{ext}^{(P_k)}\bigr),
	\label{eq:gmrt_select}
\end{equation}
under lexicographic ordering. GMR-T inherits GMR's
$O(K|E|\log|V|)$ per-event Yen runtime and adds only an
$O(K)$ post-feasibility comparison. When no transition-cheaper
alternative exists in $\mathcal{F}$, GMR-T degenerates exactly to
GMR; when one does, GMR-T captures the reduction without an explicit
detour penalty---its candidates are confined to GMR's $K$-best set,
though the selected lower-transition candidate may be marginally longer
than the GMR-optimal one.

\subsection{BD-TPAR: bounded-detour TPAR}\label{sec:bd_tpar}
BD-TPAR runs the TPAR augmented-graph candidate generator
(Section~\ref{sec:tpar}) but enforces a hard cap on the working
path's physical length. Let $L^\star(s,d)$ denote the length of the
shortest physical $s\!\to\!d$ path (a single Dijkstra computation
on physical lengths). For each augmented candidate $P_k$, BD-TPAR
accepts $P_k$ only if
\begin{equation}
	L(P_k) \le \delta\,L^\star(s,d),
	\label{eq:bd_tpar_cap}
\end{equation}
with $\delta\ge 1.0$ a tunable parameter (we use $\delta=1.2$ as
default, i.e., a working path may detour at most 20\% past the
shortest physical path to avoid a transition). If no
TPAR-augmented candidate fits the cap, BD-TPAR falls back to the
GMR candidate set, so its coverage is never strictly worse than
GMR's. At $\delta=1.0$ BD-TPAR collapses to a GMR-like scheme; at
$\delta=\infty$ it collapses to TPAR; in between, $\delta$ is an
explicit, operator-tunable detour budget.

\subsection{TPAR: transition-penalty-aware routing}\label{sec:tpar}
TPAR makes the per-transition penalty part of the path-selection
objective itself. We build a {state-augmented graph} on the
Cartesian product of the physical node set and the fiber-type set
$\{\text{HCF}, \text{SMF}\}$: each physical node $v$ becomes a pair
of augmented nodes $(v, \text{HCF})$ and $(v, \text{SMF})$
distinguished by the incoming fiber type on the edge that reached
them. An augmented edge from $(u, f_\text{in})$ to
$(v, f_\text{out})$ exists if and only if a physical link $u\to v$ of fiber
type $f_\text{out}$ exists, with cost
\begin{equation}
	w^\text{TPAR}_{(u,f_\text{in})\to(v,f_\text{out})}
	\;=\; \frac{L_{uv}}{L_\text{max}} \;+\; P[f_\text{in}, f_\text{out}],
	\label{eq:tpar_weight}
\end{equation}
where $L_{uv}$ is the link length normalized by the topology
diameter and $P$ is a $2\!\times\!2$ transition-penalty matrix
whose diagonal is zero and whose off-diagonal entries encode the
relative cost of HCF$\to$SMF vs.\ SMF$\to$HCF crossings
(Section~\ref{sec:sens}). A K-shortest-paths search on the augmented
graph---implemented by iterative single-edge exclusion rather than
strict Yen spur-path bookkeeping, which yields the same $K$
physically-distinct paths in ascending cost---returns candidate paths
whose total cost contains both a length term and an explicit transition
count. This state augmentation extends the K-shortest-paths
framework~\cite{yen1971finding} with the type-of-incoming-fiber state.

\subsection{GFJ: joint GSNR-plus-transition routing}\label{sec:gfj}
GFJ replaces TPAR's additive length+penalty form with an additive
(unnormalized) combination of inverse-GSNR cost and transition cost
on the same augmented graph:
\begin{equation}
	w^\text{GFJ}_{(u,f_\text{in})\to(v,f_\text{out})}
	\;=\; \frac{1}{\mathrm{GSNR}_\text{ext}^{(uv)}}
	+ \lambda_\text{GFJ}\, P[f_\text{in}, f_\text{out}],
	\label{eq:gfj_weight}
\end{equation}
with a single tunable scalar $\lambda_\text{GFJ}\in[0,1]$. At
$\lambda_\text{GFJ}=0$ GFJ reduces to GMR on the augmented
graph; increasing $\lambda_\text{GFJ}$ adds a transition-penalty term
of growing weight on top of the (always-retained) inverse-GSNR cost.
Because the inverse-GSNR term ($\mathcal{O}(10^{-2})$ per edge) and
the penalty term ($\mathcal{O}(1)$ per transition) are on very
different numerical scales, even a small $\lambda_\text{GFJ}$ makes
the penalty dominate; the practical consequence of this is the
sharp-threshold behavior: even small $\lambda_\text{GFJ}$ values drive blocking and carried-traffic performance to saturate near the TPAR level, leaving little usable middle ground. We use $\lambda_\text{GFJ}=0.3$ throughout, placing the scheme firmly in the saturated regime.\footnote{The acronym
	``GFJ'' is the name we use in this work for the
	{G}SNR/{F}iber-transition {J}oint formulation; it is
	not standard terminology.}

\subsection{Unified provisioning loop}\label{sec:loop}
For each incoming demand the provisioning loop produces up to
$K=5$ candidate working-paths from the scheme-specific router along
with a link-disjoint protection path. The procedure for each
candidate is: (i)~enforce the latency-asymmetry hard constraint
$|\tau_\text{work}-\tau_\text{prot}| \le \Delta\tau_\mathrm{max}$
(HC1); (ii)~apply the Extended-GSNR feasibility check of
\eqref{eq:eta_trans} with a 1~dB margin on both the working and
the protection path (HC2); (iii)~find the lowest-index wavelength
that is free on every link of both paths, using a vectorized
bitmap OR (HC3); (iv)~commit the wavelength and release it on the
demand's departure event. Blocking is recorded by the first gate
each rejected demand fails. In the current router a demand can fail
at one of four points -- latency-asymmetry (HC1), GSNR-infeasibility
(HC2), no common wavelength (HC3), or no link-disjoint
working+protection pair (``no-working-path'') -- and these four rates
sum exactly to the overall blocking probability.
\begin{figure*}[t]
	\centering
	\includegraphics[width=0.73\textwidth]{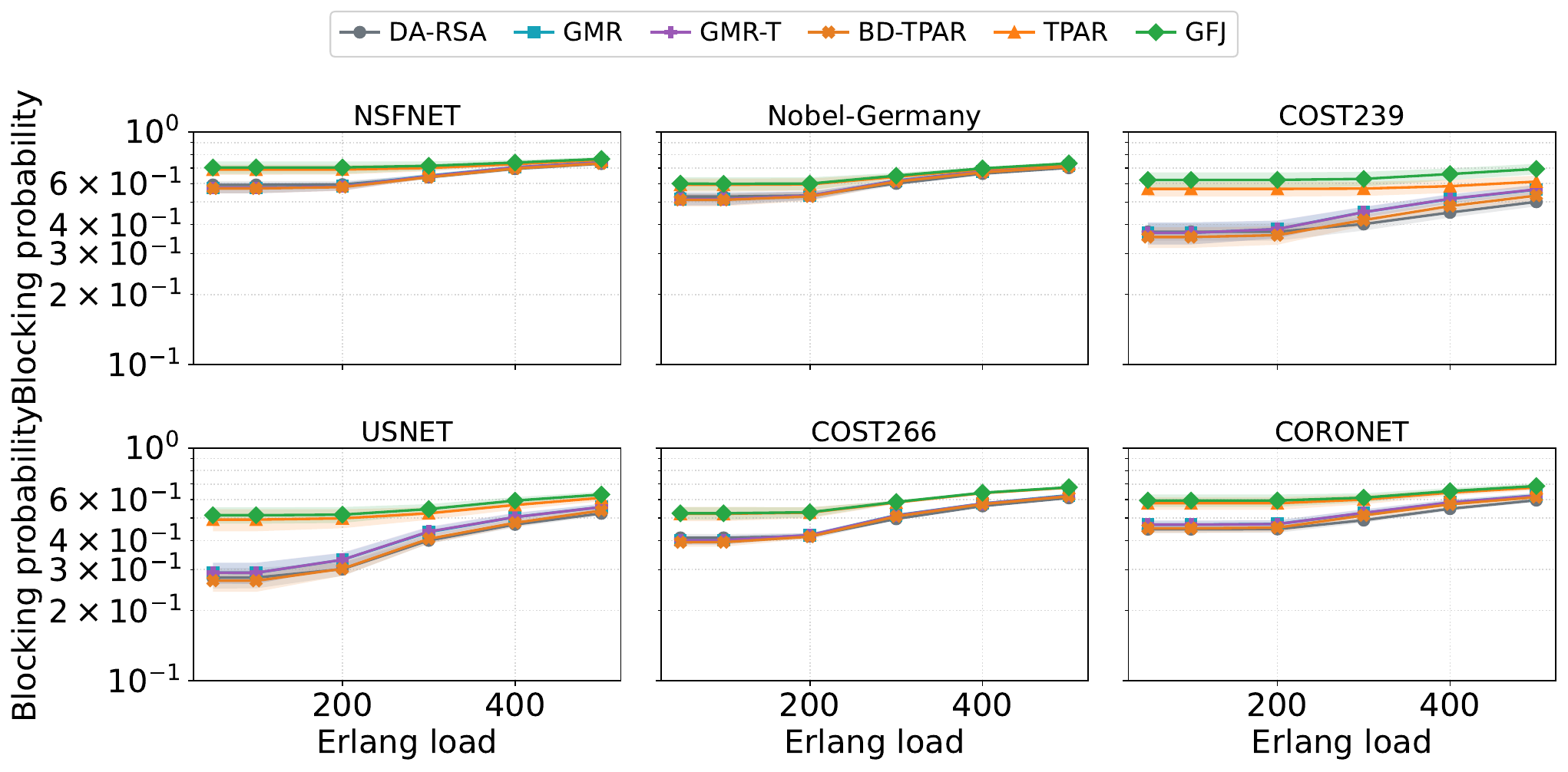}
	\caption{Blocking probability vs.\ Erlang load at 50\% HCF; bands
		are 95\% confidence intervals over ten seeds.}
	\label{fig:bp_load}
	\vspace{-.4cm}
\end{figure*}
\section{Parameter Settings}\label{sec:sens}

Section~\ref{sec:results_c} uses one fixed setting per free parameter,
listed in Table~\ref{tab:params}. The transition-penalty matrix
encodes the asymmetric per-direction operational cost of fiber-type
crossings: HCF$\to$SMF crossings cost more than SMF$\to$HCF crossings
because the EDFA on the SMF side following the splice must lift its
gain to cover both the splice loss and the higher SMF span loss,
producing a larger gain-transient excursion. The per-transition penalty
$\eta_\text{trans}$ is swept in Section~\ref{sec:eta_sweep}; all other
parameters are held fixed.

\begin{table}[h]
	\centering
	\small
	\begin{tabular}{lc}
		\toprule
		Parameter & Value \\
		\midrule
		Candidate-path count $K$ (Yen's K-SP)            & 5 \\
		GFJ transition weight $\lambda_\text{GFJ}$       & 0.3 \\
		BD-TPAR detour cap $\delta$                      & 1.2 \\
		Per-transition GSNR penalty $\eta_\text{trans}$  & \SI{0.4}{\decibel} \\
		Engineering margin $m$                           & \SI{1.0}{\decibel} \\
		Transition matrix (H$\to$S, S$\to$H)             & 1.00, 0.89 \\
		Quasi-dynamic weight-table dead-band             & 4--6 channels \\
		Latency-asymmetry limit $\Delta\tau_\text{max}$  & \eqref{eq:dtmax} \\
		Cable failure rate $\lambda_\text{cab}$          & \SI{1e-4}{\per\hour} \\
		Splice failure rate $\lambda_\text{spl}$         & \SI{1e-5}{\per\hour} \\
		Cable MTTR $\mathrm{MTTR}_\text{cab}$            & \SI{12}{\hour} \\
		Splice MTTR $\mathrm{MTTR}_\text{spl}$           & \SI{6}{\hour} \\
		\bottomrule
	\end{tabular}
	\caption{Fixed simulation parameters.}
	\label{tab:params}
\end{table}
\vspace{-.8cm}
\section{Results --- C Band}\label{sec:results_c}
\vspace{-.15cm}
All numerical results in this section are drawn from 10\,800 trials
(6~topologies $\times$ 5~HCF fractions $\times$ 6~loads $\times$
10~seeds $\times$ 6~schemes) with 3\,000 events per trial (600 warm-up,
2\,400 measurement), plus two dedicated sensitivity sweeps (the
$\eta_\text{trans}$ sweep and the contiguous-rollout comparison) run at
the same headline operating point of 50\% HCF and 300~Erlang. The
per-transition GSNR penalty $\eta_\text{trans}=\SI{0.4}{\decibel}$ and
the BD-TPAR detour cap $\delta=1.2$ are active throughout. Unless
noted, a scheme's carried-traffic lift is the mean over the six
topologies of its per-topology percentage lift relative to DA-RSA; pooling demands across topologies
instead shifts the figures by 1--2 points but not the ranking. The
section runs from headline results to mechanism, ending with the
L-band and availability cross-checks.
\subsection{Blocking probability vs.\ load}\label{sec:bp_vs_load}
Fig.~\ref{fig:bp_load} plots blocking probability against Erlang load
at 50\% HCF. The six schemes split into three behaviors. The fiber-blind
pair (DA-RSA, GMR) gives the lowest blocking baseline; under heavy
load (300~Erlang) DA-RSA achieves the minimum point-estimate blocking
across all six topologies (0.40 on COST239 up to 0.64 on NSFNET),
because its shortest physical paths claim
wavelengths most efficiently when spectrum is tight. The middle-ground
group (GMR-T, BD-TPAR) tracks DA-RSA closely---their reranking and
bounded detour keep path lengths near the shortest feasible option.
The transition-minimizers (TPAR, GFJ) block substantially more on
every topology because their path selection, constrained to avoid
fiber-type crossings, restricts route diversity and raises the
latency-asymmetry rejection rate (HC1; Section~\ref{sec:block_reasons}). The per-topology differences among
the four leading schemes are modest at 95\% confidence; the robust
effects are TPAR/GFJ's 20--25\% carried-traffic loss and the
transition-count reductions, not the small blocking gaps.

\paragraph{Why a plain GSNR objective is not enough---and what it
pays.} Before crediting the transition-aware schemes, note what a
plain GSNR objective achieves on its own---the baseline any explicit
scheme must beat. On hybrid topologies the inverse-Extended-GSNR
weight favors HCF-heavy paths
($\gamma_\mathrm{HCF}\!\ll\!\gamma_\mathrm{SMF}$), which do tend to
cross fewer fiber-type boundaries---but only as a weak,
topology-dependent side effect. Averaged over the six topologies GMR
trims $N_\text{trans}$ by just $\sim$4\% relative to DA-RSA
(Fig.~\ref{fig:mod_lift}(b)), and on some it crosses {more} boundaries than DA-RSA: a GSNR-optimal path
is not systematically transition-light. It also pays in spectrum: under
wavelength continuity an accepted demand occupies one wavelength on
{every} link of both its working and protection paths, and GMR's
GSNR-optimal routes are on average longer than DA-RSA's shortest
routes, so each acceptance consumes more link-wavelength resources,
fewer demands fit concurrently, and carried traffic drops by
$\sim$4\%. This is precisely why an
{explicit} transition term is needed: every scheme further along
the progression GMR $\to$ GMR-T $\to$ BD-TPAR $\to$ TPAR $\to$ GFJ
trades more of the same currency---spectrum occupied by longer
routes---for a reliable reduction in transitions.

Two observations: carried traffic saturates past a mid-load knee, so the
blocking and carried-traffic orderings agree; and latency asymmetry
does not differentiate {accepted} demands because $\Delta\tau_\text{max}$ clips the tails
identically---it matters instead as a {blocking} mechanism
(Section~\ref{sec:block_reasons}).

\subsection{Modulation mix and carried-traffic cost}\label{sec:throughput}
\begin{figure*}[t]
	\centering
	\includegraphics[width=.7\textwidth]{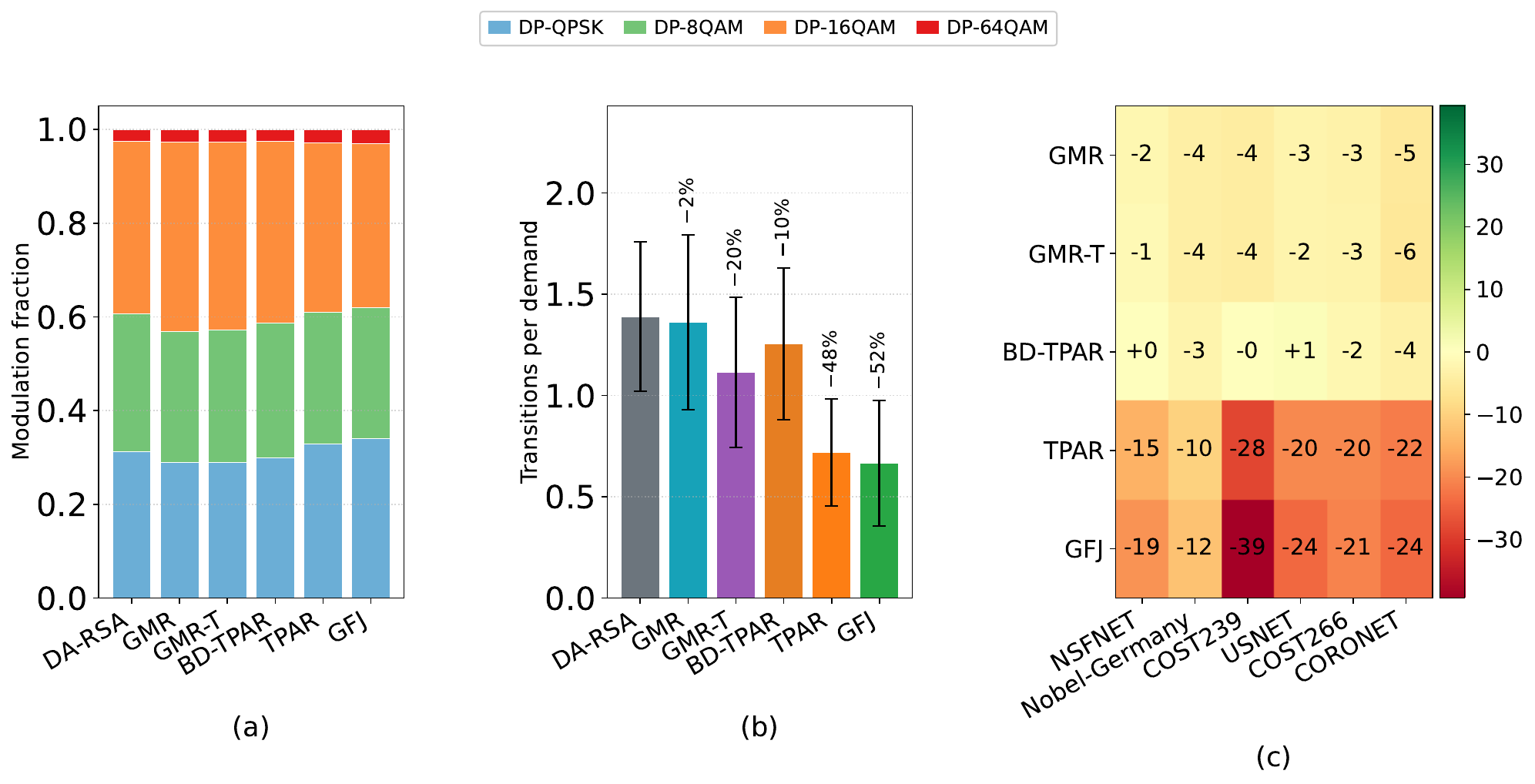}
	\vspace{-.4cm}
	\caption{The transition/throughput trade-off at 50\% HCF,
		300~Erlang. (a)~Modulation mix per scheme (mean over six
		topologies). (b)~Cross-fiber transitions per accepted demand
		(bars: six-topology mean; whiskers: spread across topologies),
		annotated with the reduction relative to DA-RSA.
		(c)~Carried-traffic lift over DA-RSA per scheme and topology.}
		\vspace{-.3cm}
	\label{fig:mod_lift}
\end{figure*}

Fig.~\ref{fig:mod_lift} presents both sides of the central
trade-off. Panel~(b) shows what each scheme buys: relative to
DA-RSA's 1.39 transitions per accepted demand, GMR-T removes
$\sim$22\%, BD-TPAR $\sim$11\%, and TPAR/GFJ $\sim$50--56\%.
Panel~(c) shows what each pays in carried traffic: BD-TPAR is the
closest non-baseline scheme at only $-1.2\%$, GMR and GMR-T sit near
$-3.5\%$, and only TPAR and GFJ separate sharply at $-19.2\%$ and
$-23.3\%$. Panel~(a) supplies the physical mechanism behind the
cost: the minimizers' longer, detoured paths drop the GSNR margin
below the DP-16QAM threshold more often, shifting demands to
lower-order formats that carry fewer bits. Taken together: BD-TPAR
buys an $\sim$11\% transition reduction for only a $\sim$1\% throughput
cost, GMR-T buys $\sim$22\% for $\sim$3\%, while TPAR/GFJ buy
$\sim$50--56\% at more than an order of magnitude higher cost.

\subsection{HCF fraction sweep}\label{sec:hcf_sweep}
\begin{figure}[t]
	\centering
	\includegraphics[width=\columnwidth]{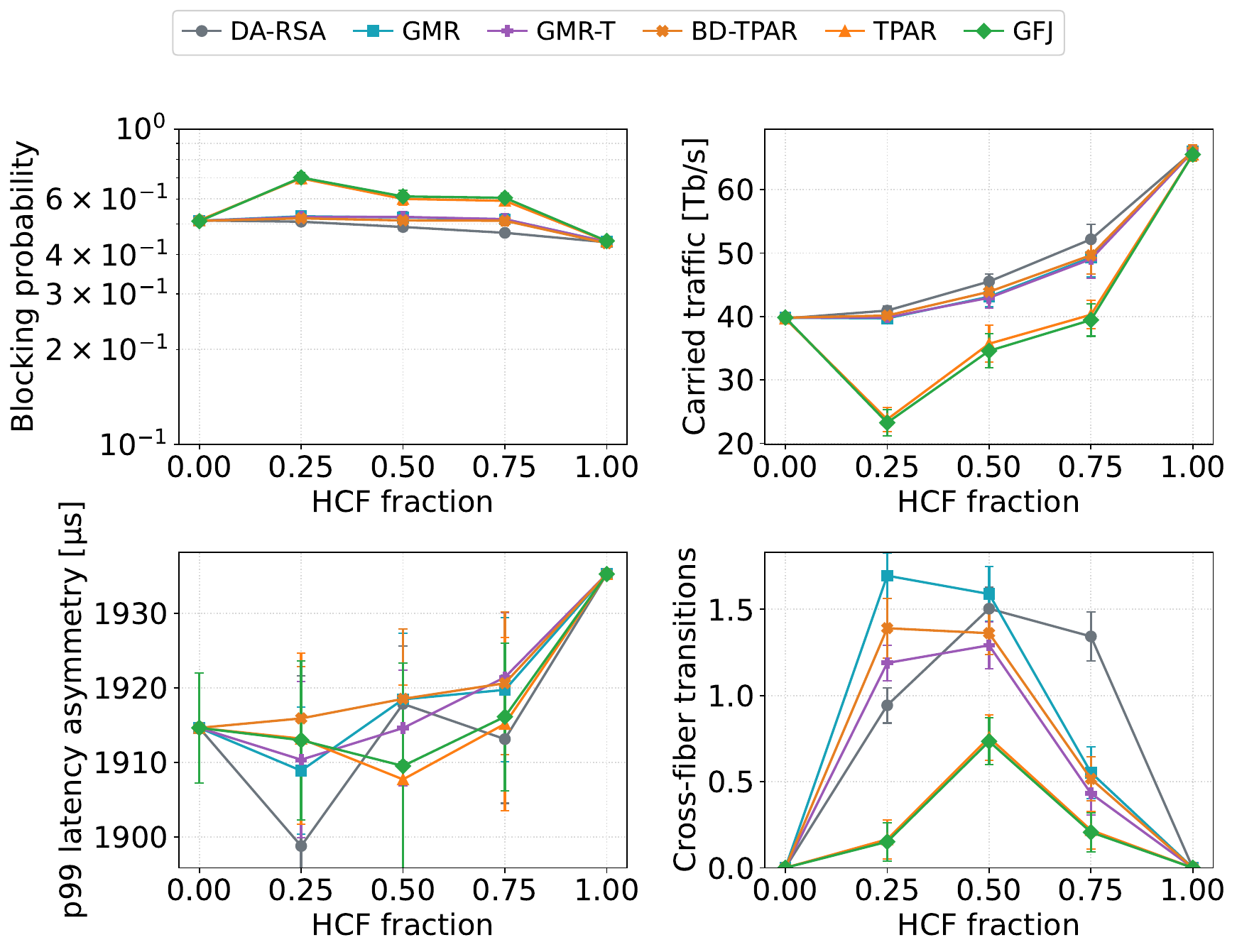}
	\caption{Blocking probability, carried traffic, p99 latency
		asymmetry, and cross-fiber transition count vs.\ HCF fraction
		(CORONET, 300~Erlang).}
		\vspace{-.3cm}
	\label{fig:hcf_sweep}
\end{figure}

Fig.~\ref{fig:hcf_sweep} shows the trade-off versus HCF fraction on
CORONET at 300~Erlang. At the boundary fractions
($p_\text{HCF}\in\{0,1\}$) the schemes are indistinguishable, since
there are no transitions to weight; the spread opens at intermediate
fractions. The transition count per accepted demand is zero at the
boundaries and peaks at 50\% HCF, where the four fiber-blind and
middle-ground schemes cluster near 1.3--1.6 (their order within the
cluster is within seed noise) while TPAR and GFJ drop to near 0.75.

\subsection{HCF deployment pattern: random vs.\ contiguous}\label{sec:rollout_compare}
\begin{figure*}[t]
	\centering
	\includegraphics[width=.59\textwidth]{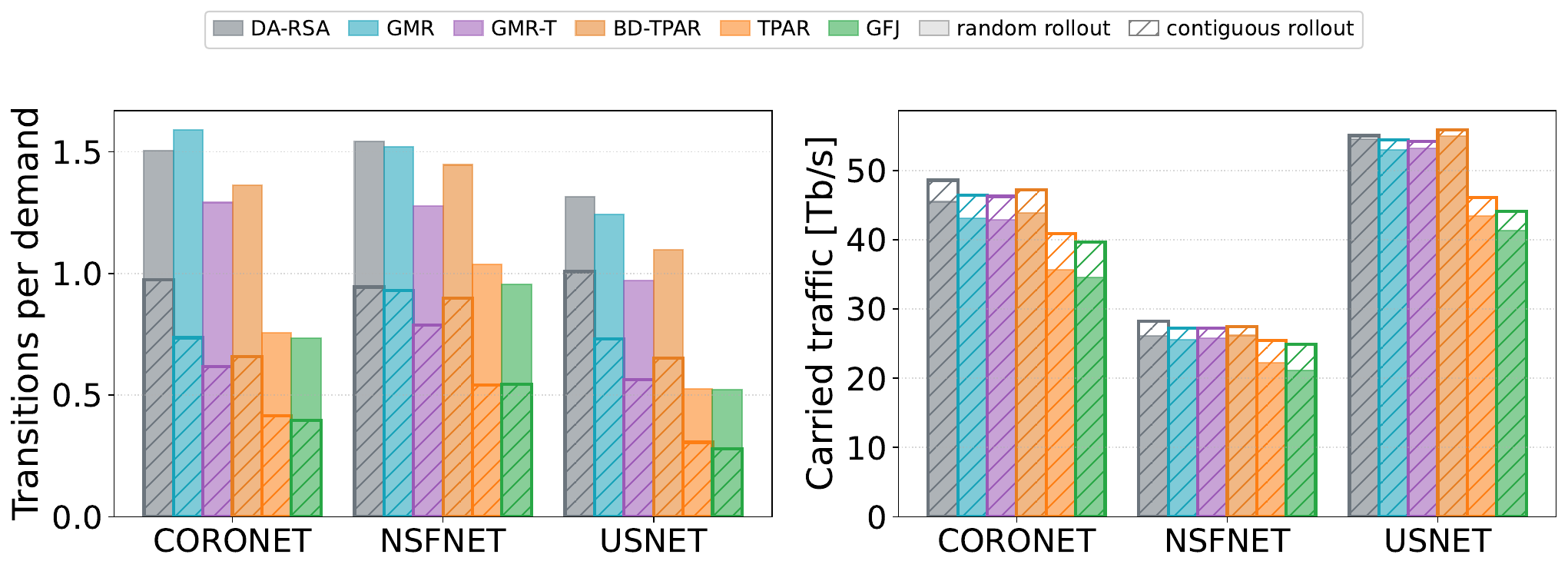}
	\vspace{-.3cm}
	\caption{Cross-fiber transitions per accepted demand (left) and
		carried traffic (right) under uniform-random (solid) vs.\
		contiguous (hatched) HCF deployment ($p_\text{HCF}=0.5$,
		300~Erlang).}
	\label{fig:rollout_compare}
	\vspace{-.3cm}
\end{figure*}

All results so far use {uniform-random} HCF assignment: the HCF
links are a uniformly-random subset of exactly $\lfloor
p_\text{HCF}|E|\rfloor$ edges, drawn without regard to adjacency, so HCF
and SMF interleave freely and per-route transition counts are high---a
strongly fragmented deployment representative of uncoordinated,
opportunistic upgrades. The alternative,
{contiguous} deployment, grows the HCF plant as a single
connected cluster by BFS expansion from a seed link
(Section~\ref{sec:topo}), modeling an operator that upgrades one
region at a time; real rollouts are likely closer to this end.
Fig.~\ref{fig:rollout_compare} compares the two on three topologies
at $p_\text{HCF}=0.5$ and 300~Erlang. Contiguous deployment cuts the
per-demand transition count on every scheme, by 23--54\% (mean
$\sim$42\%), and raises carried traffic most where fragmentation hurt
most and for the schemes that detour most to avoid transitions---on
NSFNET, GFJ gains $\sim$18\% and TPAR $\sim$14\%, with three-topology
mean gains of $\sim$13\% (GFJ) and $\sim$12\% (TPAR) versus only
$\sim$5\% for the leading schemes (DA-RSA, GMR, GMR-T, BD-TPAR).
Deployment pattern is thus a throughput-preserving lever reaching
transition reductions comparable to TPAR/GFJ's without their penalty;
the transition-minimizers' advantage shrinks under contiguous rollout,
so the decision rule applies most strongly on the fragmented end.

\subsection{Per-transition penalty sensitivity}\label{sec:eta_sweep}
\begin{figure}[t]
	\centering
	\includegraphics[width=0.93\columnwidth]{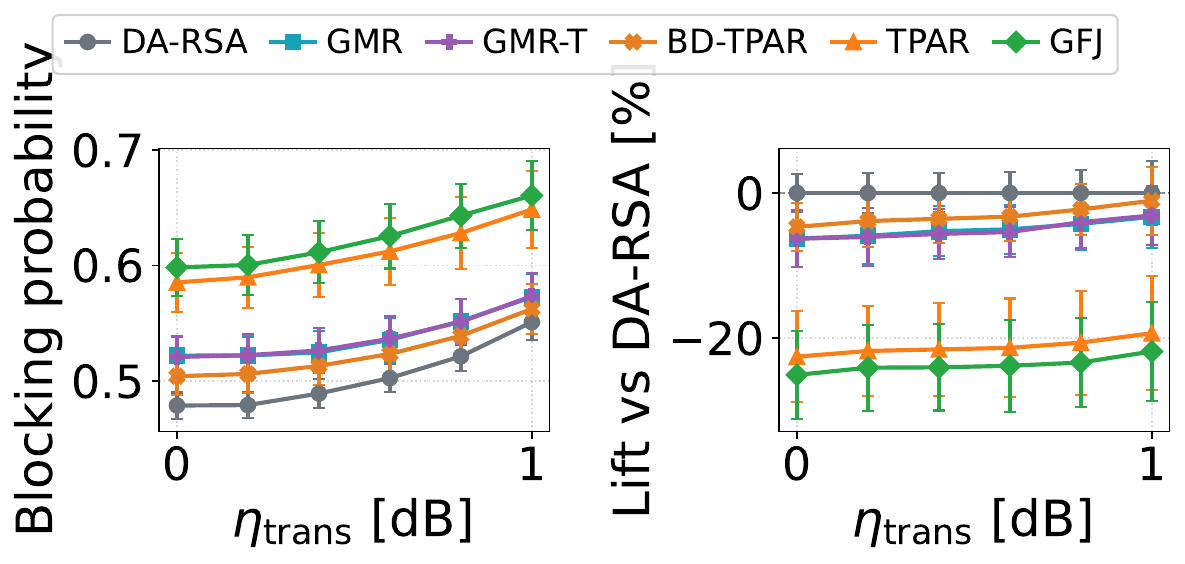}
	\vspace{-.3cm}
	\caption{Blocking probability (left) and carried-traffic lift over
		DA-RSA (right) vs.\ the per-transition penalty
		$\eta_\text{trans}$ (CORONET, 50\% HCF, 300~Erlang).}
	\label{fig:eta_sweep}
	\vspace{-.3cm}
\end{figure}

Since $\eta_\text{trans}$ is a calibrated parameter
(Section~\ref{sec:eta_trans}), the natural question is whether the
conclusions depend on its value. Fig.~\ref{fig:eta_sweep} sweeps it
over $[0,1]$~dB on CORONET at 50\% HCF and 300~Erlang (360 dedicated
trials); they do not. Blocking rises with $\eta_\text{trans}$ as
more candidates cross the modulation cliff---monotonically for every
scheme, with DA-RSA nearly flat below $0.2$~dB (0.479 at both $0$ and
$0.2$~dB, where the penalty is still too small to prune many
candidates) before climbing to 0.551 at $1$~dB. The ordering is stable
across the range: BD-TPAR tracks 2--3 points above DA-RSA and slightly
below GMR/GMR-T, its lift over DA-RSA rising from $-5\%$ toward $-1\%$
as $\eta$ grows (its fewer transitions worth more as each costs more),
while TPAR/GFJ stay in a $-19$ to $-25\%$ band.

\subsection{Block-reason decomposition}\label{sec:block_reasons}
\begin{figure}[t]
	\centering
	\includegraphics[width=0.99\columnwidth]{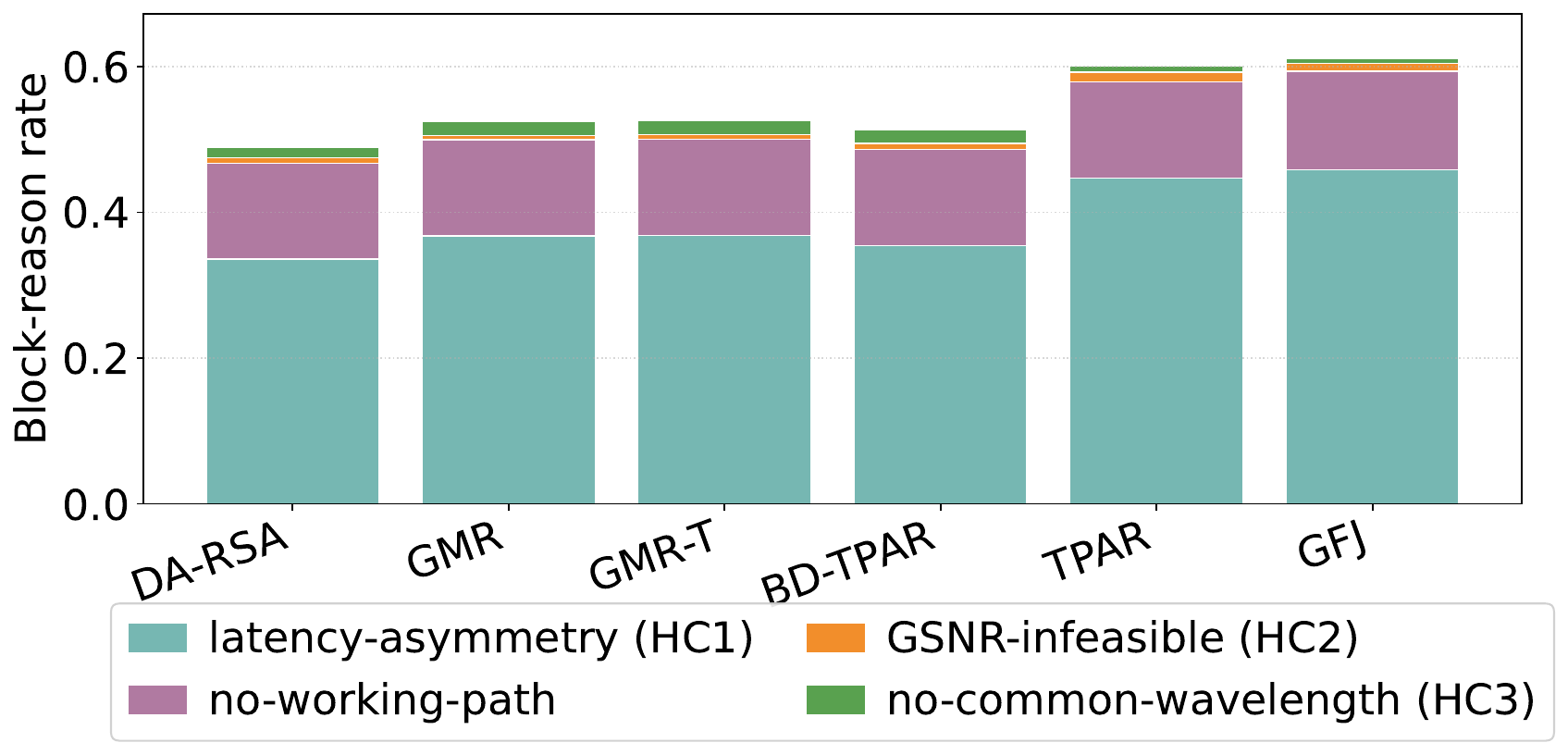}
	\caption{Block-reason rates per scheme (CORONET, 50\% HCF,
		300~Erlang).}
	\label{fig:block_reasons}
	\vspace{-.3cm}
\end{figure}

Fig.~\ref{fig:block_reasons} answers {why} demands are blocked.
Blocking decomposes into four classes summing to the total:
latency-asymmetry (HC1---the receiver differential-delay bound of
\eqref{eq:dtmax}, Section~\ref{sec:mod}), no-working-path,
GSNR-infeasibility (HC2), and no-common-wavelength (HC3), each demand
attributed to the gate at which its preferred route first failed.
At 50\% HCF / 300~Erlang on CORONET, HC1 dominates
every scheme (69--75\%),
followed by no-working-path (22--27\%); spectrum-related rejections
are minor (HC3 1--4\%, HC2 $\sim$1--2\%). Demands are lost mainly
because no working/backup pair with matched HCF/SMF delays
exists---not because wavelengths run out. This also explains the
minimizers' larger traffic loss: avoiding HCF--SMF crossings
restricts route diversity and magnifies the delay mismatch, pushing
their HC1 share to 74--75\% versus $\sim$69\% for the other four.

\subsection{L-band cut with per-channel CO\textsubscript{2} absorption}\label{sec:lband}

CO\textsubscript{2} absorption in HCF-guided L-band light arises from
gas-phase rovibrational lines of the $2\nu_1{+}\nu_3$ combination band
(origin $\approx190.22$~THz), spaced $\approx46.4$~GHz apart, each
$\approx1$~GHz wide and reaching $\approx0.1$~dB/km at the strongest
lines~\cite{chen2025co2}. Our 80-channel, 75-GHz L-band grid spans
184.783--190.708~THz (1622--1572~nm). Because 75~GHz is not a multiple
of the line spacing, channel centers fall at different offsets from the
measured absorption lines~\cite{chen2025co2,photonics13060559}. However,
a 1-GHz line overlaps only a thin portion of a 64-GBd channel's
$\approx70$-GHz occupied bandwidth, so the band-averaged excess loss is
far below the line-center peak. Even channel~78, whose center nearly
coincides with a strong R-branch line and would incur
$\approx7.4$~dB/100~km under point sampling, experiences only
$\approx0.4$~dB/100~km after band averaging. Most channels experience
far less, while channels 0--17 occupy the CO\textsubscript{2}-quiet
region where high-$J$ lines are Boltzmann-suppressed.

We represent this effect using per-channel GSNR look-up tables, assigning
each wavelength slot on every HCF link the effective loss
\[
\alpha_\text{eff}(\nu_k)
=\alpha_\text{HCF}
+\bar{\alpha}_{\mathrm{CO}_2}(\nu_k).
\]
Here, $\bar{\alpha}_{\mathrm{CO}_2}(\nu_k)$ is a phenomenological comb of
1-GHz-FWHM Lorentzians whose positions and amplitudes are fitted to
the measured L-band spectrum of~\cite{chen2025co2} and integrated over
the 64-GBd signal bandwidth rather than sampled at the carrier. This is
a first-order OSNR treatment: the EDFA compensates the band-averaged
excess loss and consequently raises ASE. It is a routing-layer penalty
map, not a first-principles line-by-line HITRAN model, and does not
capture narrow-band spectral distortion within a channel. GSNR is
precomputed for three occupancy states, as in the C-band analysis.
Because the profile is non-monotone in channel index, the penalty seen
by first-fit assignment depends on load and path length as well as the
selected channel.

\begin{figure*}[t]
	\centering
	\includegraphics[width=0.78\textwidth]{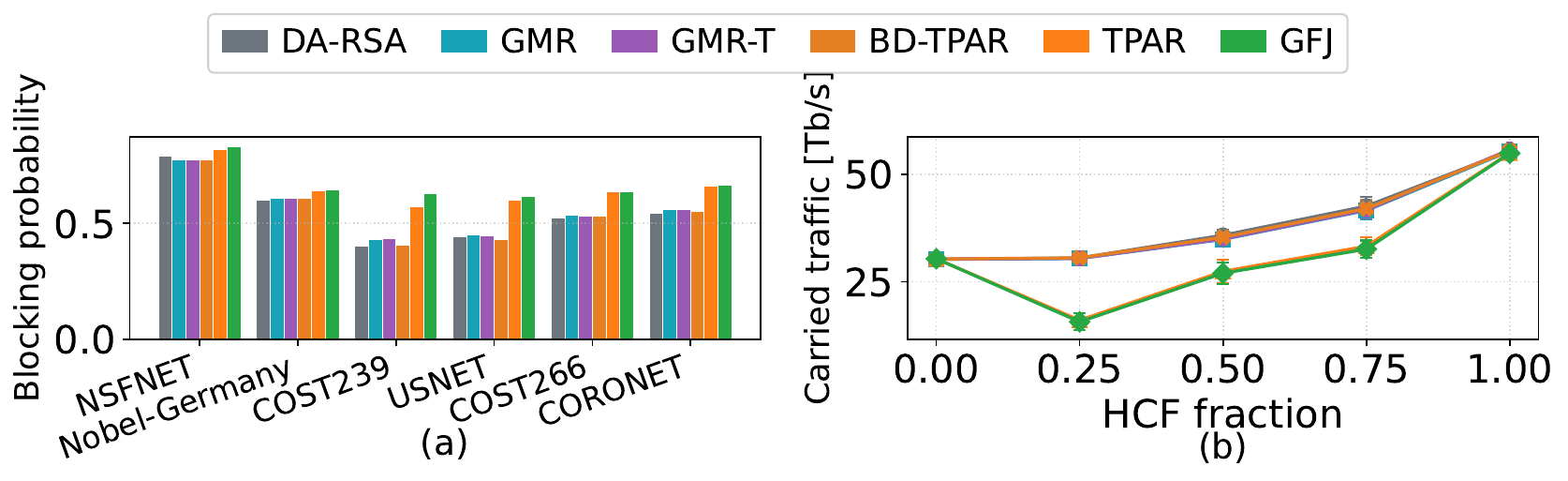}
	\vspace{-.4cm}
	\caption{L-band results with per-channel CO\textsubscript{2}
		penalties. (a)~Blocking per topology at 50\% HCF and
		300~Erlang; the scheme ordering matches the C-band.
		(b)~Carried traffic versus HCF fraction on CORONET; traffic
		increases with HCF deployment.}
	\label{fig:lband_summary}
	\vspace{-.3cm}
\end{figure*}

We ran 10{,}800 trials
(6 topologies $\times$ 5 HCF fractions $\times$ 6 loads $\times$
10 seeds $\times$ 6 schemes) on the 80-channel L-band grid.
Fig.~\ref{fig:lband_summary}(a) shows that the C-band ordering is
preserved at 50\% HCF and 300~Erlang: DA-RSA has the lowest blocking on
every topology, ranging from 0.40 on COST239 to 0.79 on NSFNET; TPAR and
GFJ incur 20--25\% mean carried-traffic penalties relative to DA-RSA;
and BD-TPAR remains within $\sim1\%$ of DA-RSA
($+1.0\%$ by the per-topology mean). The small band-averaged
CO\textsubscript{2} penalty leaves absolute L-band capacity comparable
to the C-band, whereas a
point-sampled treatment would instead charge each carrier the full
line-center absorption.

Fig.~\ref{fig:lband_summary}(b) likewise shows that CORONET carried
traffic increases monotonically with HCF fraction, from $\sim30$~Tb/s
at the all-SMF endpoint to $\sim56$~Tb/s at all-HCF for BD-TPAR. The
band-averaged gas penalty is therefore too small to offset HCF's lower
loss and near-zero Kerr nonlinearity, making L-band HCF deployment
net-beneficial. The residual penalty is confined to channels whose
$\sim70$-GHz bands overlap dense line clusters. Thus, both the
scheme-choice rule and deployment incentive are band-invariant: the
scheme ranking is unchanged, and greater HCF deployment helps in both
bands.

If the residual penalty must be removed, its known spectral locations
make it a spectrum-assignment problem rather than a routing problem:
filling quiet channels 0--17 first and line-adjacent channels such as
78 last would recover the remaining few percent without changing path
selection. Pre-emphasis and spectral-avoidance capacity
maps~\cite{photonics13060559} provide the required per-channel margin
bookkeeping, while DSP mitigation using pre-characterized absorption
profiles is emerging~\cite{wang2025co2,chen2025co2}. 
  \begin{figure}[t]
 	\centering
 	\includegraphics[width=\columnwidth]{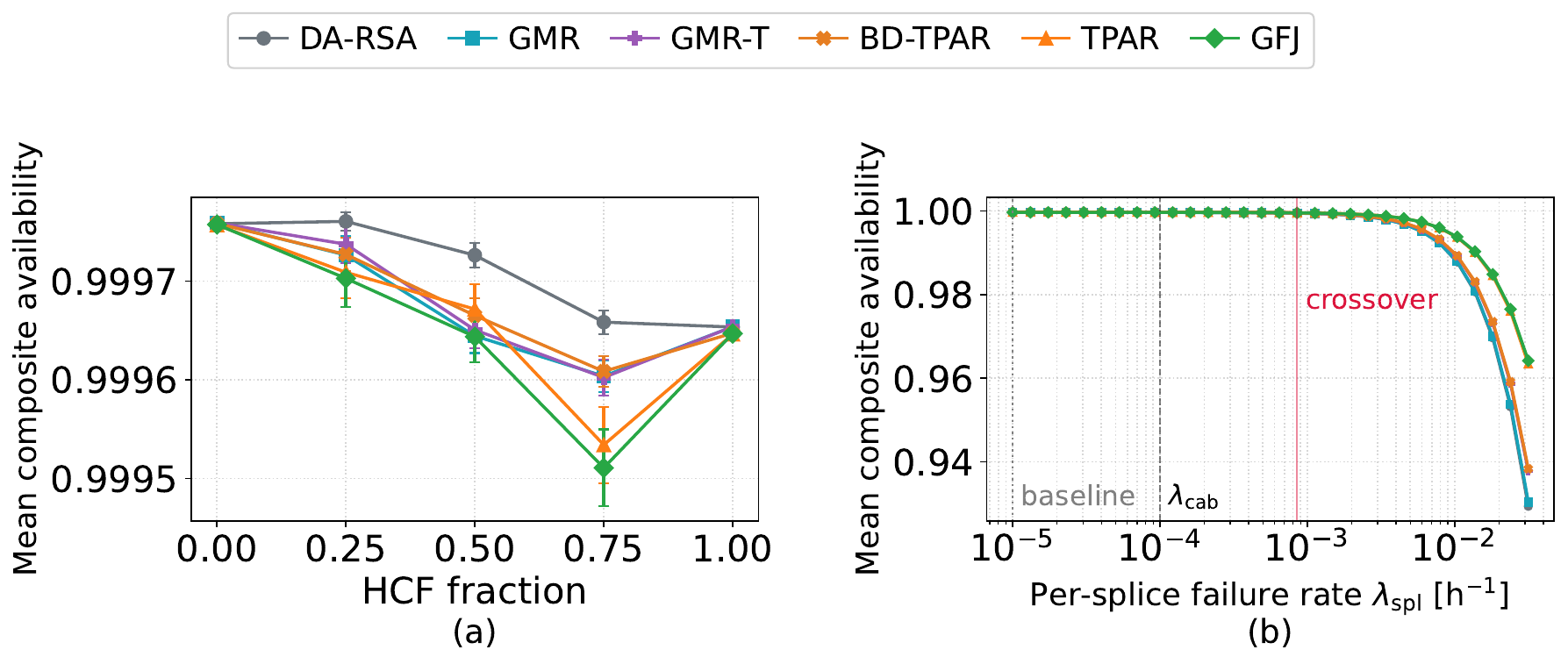}
 	\caption{Composite 1+1 availability on CORONET at 300~Erlang.
 		(a)~Mean composite availability versus HCF deployment fraction. (b)~Mean composite availability versus the per-splice
 		failure rate at 50\% HCF, recomputed over the captured paths of
 		accepted demands without rerouting. The gray dotted and black dashed
 		lines indicate the baseline splice-failure rate and cable-failure rate, respectively. The red
 		line marks the first crossover, where TPAR
 		first exceeds DA-RSA in mean composite availability.}
 	\label{fig:composite_availability_combo}
 \end{figure}
\subsection{Composite availability with per-splice failure}\label{sec:results_avail}
Does transition reduction buy reliability? At our calibration, no.
Extra transitions add outage exposure, but composite availability is
dominated by the cable-cut and per-channel outage terms, both scaling
with working-path length, so the shortest-path scheme (DA-RSA) is the
most available by a small margin; all six lie between 0.9996 and 0.9997
at $p_\text{HCF}=0.5$ (Fig. \ref{fig:composite_availability_combo}(a)), none reaching
five-nines because the residual per-channel and cable-cut terms cap the
1+1 composite there.
Sweeping the per-splice rate over each demand's captured paths,
transition-aware schemes begin to overtake DA-RSA only once
$\lambda_\text{spl}$ reaches approximately
$8.5\times10^{-4}$~h$^{-1}$ (Fig. \ref{fig:composite_availability_combo}(b)); TPAR is the first scheme to cross at this
point. This rate is approximately $85\times$ the already-conservative
baseline and implies a splice mean time between failures of about
1200~h, still far above plausible fusion-splice failure rates.
Availability is therefore a caution against over-claiming reliability
benefits, not an argument for transition-aware routing; the case for
transition awareness rests on the operational costs discussed in
Section~\ref{sec:transition_cost}.

\section{Discussion}\label{sec:discussion}

\subsection{Why a transition has an external cost}\label{sec:transition_cost}
We define $C_\text{ext}$ as the operator's {per-unit external
cost} of an HCF$\leftrightarrow$SMF transition, in carried-traffic
units. It aggregates three operational-incident classes invisible to
blocking and throughput. (i)~Each interface causes an EDFA gain-target
discontinuity: surviving channels see millisecond-scale, multi-decibel
transients in cascaded
amplifiers~\cite{srivastava1997edfa,tancevski1999swings}, consuming
optical-signal-to-noise-ratio (OSNR) margin and triggering ROADM
retunes. (ii)~Each splice is a discrete failure point (rate
$\lambda_\text{spl}$), so lifetime outage exposure scales with the
transition count. (iii)~Working and protection paths of different
fiber type present different accumulated CD, so hitless switchover
requires the transponder DSP to hold equalizer state for
both~\cite{tang2024multilayer} (no public quantification of this
specific cost is known to us). We model (i) via $\eta_\text{trans}$
and (ii) via $\lambda_\text{spl}$; (iii) is qualitative only. A high
$C_\text{ext}$ reflects an operator whose EDFA-retune budget, MTTR
service-level agreements (SLAs), and transponder overhead price
transitions heavily; a low $C_\text{ext}$, one constrained only by
spectrum.

\subsection{Decision rule}\label{sec:decision_rule}
The rule turns the simulation outputs into a scheme choice in three
steps. \emph{Step 1: read two numbers per scheme} off
Fig.~\ref{fig:mod_lift} (50\% HCF, 300~Erlang, six-topology means): the
carried-traffic lift over DA-RSA $w_s$ (panel c) and the mean
transitions per accepted demand $\bar{n}_s$ (panel b). \emph{Step 2:
set what one transition is worth}---the operator's exchange rate
$\kappa$ (Section~\ref{sec:transition_cost}), the percent of carried
traffic it would give up to remove one transition per demand ($\kappa=0$
with no transition costs, large where interface-driven maintenance
dominates).

\emph{Step 3: pick the scheme on the upper envelope.} A scheme's utility is its
throughput lift plus the value of the transitions it saves,
\begin{equation}
	U_s \;=\; w_s \;+\; \kappa\,\bigl(\bar{n}_\text{DA-RSA}-\bar{n}_s\bigr).
	\label{eq:utility}
\end{equation}
Because \eqref{eq:utility} is linear in $\kappa$, the best scheme at
any $\kappa$ is whichever utility line is highest---the {upper
envelope} of the six lines, read directly off the $(w_s,\bar{n}_s)$
values of Fig.~\ref{fig:mod_lift}. Taking the envelope over all six
(not chaining pairwise comparisons) matters: GMR never reaches it, since
GMR-T dominates GMR---more transitions saved ($\bar{n}=1.11$ vs.\ $1.36$)
at similar cost---so GMR is never optimal. The remaining schemes
partition the $\kappa$ axis into five regions, delimited by the
$\kappa$ at which adjacent utility lines cross:
\begin{itemize}
	\item $\kappa < 9$: transitions too cheap to justify any
	detour -- {use DA-RSA}.
	\vspace{-.3cm}
	\item $9 \le \kappa < 16$: {use BD-TPAR} ($\delta=1.2$; GMR
	fallback) -- $\sim$11\% fewer transitions at almost no throughput cost.
		\vspace{-.3cm}
	\item $16 \le \kappa < 40$: {use GMR-T} -- a larger $\sim$22\%
	transition cut for $\sim$3\% throughput.
		\vspace{-.3cm}
	\item $40 \le \kappa < 77$: {use TPAR}.
		\vspace{-.3cm}
	\item $\kappa \ge 77$: {use GFJ}.
\end{itemize}
Concretely: an operator willing to give up 1\% of carried traffic to
remove 0.1 transitions per demand sits at $\kappa=1/0.1=10$ (BD-TPAR);
one willing to give up 3\% for the same 0.1 sits at $\kappa=30$ (GMR-T).

Two qualifiers bracket the rule. First, it matters most under
fragmented rollout. The exchange rate $\kappa$ is a property of the
operator's cost structure, not the topology; what contiguous deployment
changes is the transition-savings term
$(\bar{n}_\text{DA-RSA}-\bar{n}_s)$ multiplying $\kappa$ in
\eqref{eq:utility}. Removing $\sim$40\% of transitions structurally
(Section~\ref{sec:rollout_compare}) shrinks that term for every scheme,
so the utility advantage of transition-aware routing collapses toward
zero even at large $\kappa$. Second, the breakpoints are heavy-load
numbers ($w_s$ at 300~Erlang). At light load (50--100~Erlang) the
middle-ground schemes {edge ahead} of DA-RSA ($+2$ to $+3\%$),
their transition-light paths keeping more GSNR margin while spectrum is
plentiful, so below the blocking knee they are safe at {any}
$\kappa$. The TPAR/GFJ penalty, however, persists at all loads ($-23$
to $-27\%$) because it stems from latency-asymmetry rejection, not
spectrum contention; the $\kappa\gtrsim40$ gate applies at any load.



\section{Conclusion}\label{sec:conclusion}

We compared six protected routing schemes on a common event-driven
simulator with an IMI-augmented GN model and a calibrated
per-transition GSNR penalty; the middle-ground schemes introduced
here---GMR-T and BD-TPAR---fill the gap between fiber-blind baselines
and aggressive minimizers. The preferred scheme depends on load
regime and operator cost structure, not on a universal ranking. Under
heavy load, latency-asymmetry rejection (HC1) binds and DA-RSA is
the throughput champion; BD-TPAR is the closest non-baseline scheme
at only $-1.2\%$ cost for $\sim$11\% fewer transitions, and GMR-T buys
$\sim$22\% fewer for $\sim$3\%. Aggressive minimizers (TPAR, GFJ) halve
transitions but cost 20--25\% of carried traffic, justified only when
the operator's transition-to-throughput exchange rate exceeds
$\kappa\approx40$ (Section~\ref{sec:decision_rule}). The ordering also holds in the L-band, where a band-averaged
CO\textsubscript{2} penalty leaves HCF net-beneficial. Contiguous HCF rollout is the strongest lever (23--54\% fewer
transitions, higher carried traffic); routing adds value on top,
most under fragmented rollout. These results replace ``which scheme
is best?'' with ``what cost structure does the operator carry?''

\begin{backmatter}
\bmsection{Funding} No funding received for this work.
\vspace{-.2cm}
\bmsection{Acknowledgment} Generative AI was used for language editing.
\vspace{-.2cm}

\bmsection{Disclosures} The authors declare no conflicts of interest.

\vspace{-.2cm}
\bmsection{Data availability} Data underlying the results presented in this paper are not publicly available at this time but may be obtained from the authors upon reasonable request.
\vspace{-.2cm}

\end{backmatter}


\bibliography{jocn_gsnr_routing}

\bibliographyfullrefs{jocn_gsnr_routing}


\end{document}